\documentstyle[twocolumn,aps,prb,epsf]{revtex}
\begin{document}
\draft
\twocolumn[\hsize\textwidth\columnwidth\hsize\csname
@twocolumnfalse\endcsname


\title{Strong-coupling theory of superconductivity 
in a degenerate Hubbard model}

\author{Tetsuya Takimoto$^1$, Takashi Hotta$^1$, and Kazuo Ueda$^{2}$}

\address{$^1$Advanced Science Research Center,
Japan Atomic Energy Research Institute, Tokai, Ibaraki 319-1195, Japan}
\address{$^2$Institute for Solid State Physics, University of Tokyo,
5-1-5 Kashiwa-no-ha, Kashiwa, Chiba 277-8581, Japan}

\date{\today}
\maketitle

\begin{abstract}
In order to discuss superconductivity in orbital degenerate systems, 
a microscopic Hamiltonian is introduced. 
Based on the degenerate model, 
a strong-coupling theory of superconductivity 
is developed within the fluctuation exchange (FLEX) approximation 
where spin and orbital fluctuations, spectra of electron, 
and superconducting gap function are self-consistently determined. 
Applying the FLEX approximation to the orbital degenerate model, 
it is shown that the $d_{x^2-y^2}$-wave 
superconducting phase is induced by increasing the orbital 
splitting energy which leads to the development and suppression 
of the spin and orbital fluctuations, respectively. 
It is proposed that the orbital splitting energy is
a controlling parameter changing from the paramagnetic
to the antiferromagnetic phase with 
the $d_{x^2-y^2}$-wave superconducting phase in between.
\end{abstract}


\vskip2pc]
\narrowtext

\section{Introduction}

Recently, 
new heavy fermion superconductors CeTIn$_5$ (T=Rh, Ir, and Co) 
have been discovered. 
These compounds have been atracted much attention, 
since varieties of ordered states are observed by changing 
transition metal ions. 
Among them, CeIrIn$_5$ and CeCoIn$_5$ are 
superconducting at ambient pressure with transition temperatures 
$T_{\rm c}$=0.4K and 2.3K, respectively.\cite{CeIrIn5,CeCoIn5}
In particular, 
CeCoIn$_5$ shows the highest superconducting transition 
temperature among Ce-based heavy fermion systems. 
On the other hand, 
CeRhIn$_5$ exhibits an antiferromagnetic 
transition at a N\'eel temperature $T_{\rm N}$=3.8K and 
becomes superconducting only under hydrostatic pressure 
larger than 15 kbar.\cite{CeRhIn5}

The important experimetal results of these materials are summarized 
as follows. 
Reflecting the fact that CeTIn$_5$ has the HoCoGa$_5$-type
tetragonal crystal structure,
quasi two-dimensional Fermi surfaces have been observed
in de Haas-van Alphen experiments of the compounds,
consistent with the band-structure calculations.\cite{Haga,Settai}
The specific heat of CeCoIn$_5$ has shown 
considerably large discontinity 
at the superconducting transition temperature 
by which the superconductivity of this compound 
is considered to be in the strong-coupling regime.\cite{CeCoIn5}
Concerning the superconducting state, nuclear relaxation rate
of CeTIn$_5$ exhibits $T^3$ behavior below $T_{\rm c}$\cite{Kohori,Zheng}
and thermal conductivity in CeCoIn$_5$ is found
to show a component with four-fold symmetry,\cite{Izawa}
which strongly suggest the $d_{x^2-y^2}$-wave pairing symmetry 
of the superconducting phase of CeTIn$_5$.
Furthermore, it has been shown that in the alloy system
CeRh$_{1-x}$Ir$_{x}$In$_5$, the superconducting phase appears
in the neighborhood of the antiferromagnetic phase.\cite{Pagliuso}
These experimental results indicate that
the CeTIn$_5$ compounds have similarity with 
high-$T_{\rm c}$ cuprates. 
Thus, it is natural to expect that the mechanism of superconductivity 
of CeTIn$_5$ is similar to 
that of high-$T_{\rm c}$ cuprates.

From the band-structure calculations of CeIrIn$_5$ and CeCoIn$_5$, 
one can see that 
characteristic features of the Fermi surfaces of these compounds, 
such as shape, volume, and the $4f$-electron weight,  
are almost the same with each other.\cite{Haga,Settai,Maehira}
On the other hand, it has been shown in theoretical studies 
of high-$T_{\rm c}$ cuprates 
that hole doping leads to the deformation of the Fermi surface 
and the structure of spin fluctuations 
around the antiferromagnetic vector is changed. 
Based on the mechanism of $d_{x^2-y^2}$-wave superconductivity 
induced by the antiferromagnetic spin fluctuation, 
variation of $T_{\rm c}$ 
in the overdoped region of high-$T_{\rm c}$ cuprates 
has been explained reasonably by this scenario.\cite{Scalapino,Moriya}
It means that hole doping 
is controlling the superconducting transition temperature 
through the deformation of the Fermi surface 
in high-$T_{\rm c}$ cuprates. 
From the point that the superconducting transition temperatures 
of CeIrIn$_5$ and CeCoIn$_5$ are quite different from each other 
in spite of the similarities of the Fermi surfaces, 
it is an improbable scenario that 
the superconductivity of CeTIn$_5$ compounds is primarily controlled 
by the carrier doping as in the high-$T_{\rm c}$ cuprates. 
Thus, other controlling 
parameters for the superconductivity should be searched in CeTIn$_5$ 
even if the superconducting mechanism 
of CeTIn$_5$ is similar to 
that of high-$T_{\rm c}$ cuprates. 
In order to find 
such a controlling parameter of superconductivity 
in the heavy fermion system CeTIn$_5$, 
theoretical study from the microscopic point of view 
is highly required.

However, theoretical studies for the heavy fermion superconductivity 
have been almost restricted in the phenomenological level \cite{GLtheory} 
because of the following problems: 
(1) It is difficult to treat the dual nature of $f$-electrons, 
coexistence of both localized and itinerant character, 
in contrast with the $d$-electron systems 
where the itinerant picture is a good starting point. 
(2) Complicated $f$-electronic states are formed 
by the combined effect of crystal structure 
and orbital degeneracy of $f$-electrons, 
which leads to multiple Fermi surfaces. 
For the first difficulty, it may be a reasonable assumption that 
even 
strongly correlated states with such a duality of $f$-electrons 
are still adiabatically continued from the state 
in weakly correlated systems which is the key assumption of 
the Fermi-liquid theory. 
With respect to the second difficulty, 
it may be appropriate to introduce a microscopic model 
considering the essential part of the complicated crystal structure. 
Furthermore, 
it may be necessary to incorporate the orbital degree of freedom 
of $f$-electrons into the model, 
since quasi-particle states are reflected by kinds of orbitals. 
Thus, in order to understand superconductivity 
in the heavy fermion system 
from a microscopic point of view, 
we should develop a microscopic theory 
based on the Fermi-liquid type theory 
using an orbital degenerate model. 
Especially, in view of the large discontinuity of 
the specific heat at $T_{\rm c}$ in CeCoIn$_5$, 
it is necessary to develop a strong-coupling theory 
for superconductivity in CeTIn$_5$ compounds. 

In this paper, we focus on the effects of orbital degrees of freedom 
on superconductivity based on a microscopic theory applied to 
a microscopic model with the orbital degeneracy. 
In the next section, 
we introduce the orbital degenerate model obtained by including 
important characters of CeTIn$_5$. 
Then, in order to study the superconducting transition 
in the orbital degenerate model, 
we develop a strong-coupling theory using the fluctuation exchange (FLEX) 
approximation \cite{FLEX} in which spin and orbital fluctuations, 
the single-particle spectrum, 
and superconducting gap function are determined self-consistently. 
Finally, we discuss experimental results for CeTIn$_5$ 
in the light of the present theory. 

\section{Model Hamiltonian}

First let us introduce local basis for $f$-electron systems. 
As is well known, 14-fold degenerate $f$-electronic states split to 
$j$=5/2 and 7/2 multiplets due to strong spin-orbit coupling 
where $j$ is total angular momentum. 
It is quite natural to consider only the lower $j$=5/2 multiplet 
contributes to low-energy excitations. 
This multiplet splits into 
$\Gamma_7$ doublet and $\Gamma_8$ quartet under cubic 
crystalline electric field (CEF), 
and the corresponding eigen-states are given by
\begin{eqnarray}
  &&|\Gamma_{7\pm} \rangle=
  \sqrt{\frac{1}{6}}|\pm \frac{5}{2}\rangle
  -\sqrt{\frac{5}{6}}|\mp \frac{3}{2}\rangle, \\
  &&|\Gamma_{8\pm}^{(1)} \rangle=
  \sqrt{\frac{5}{6}}|\pm \frac{5}{2}\rangle
  +\sqrt{\frac{1}{6}}|\mp \frac{3}{2}\rangle, \\
  &&|\Gamma_{8\pm}^{(2)}\rangle=|\pm \frac{1}{2}\rangle,
\end{eqnarray}
where $|j_{z} \rangle$ are basis of $j$=5/2 multiplet, 
and $+$($-$) in the subscripts denotes ``pseudo-spin'' up(down)
in each Kramers doublet.
Furthermore, under tetragonal CEF, two $\Gamma_7$ and one $\Gamma_6$ 
Kramers doublets are formed. 
Experimental data of magnetic susceptibility of 
CeTIn$_5$ analysed by using the CEF theory seem to be consistent 
with the level scheme where the two $\Gamma_7$ are lower 
than the $\Gamma_6$. \cite{Takeuchi,Shishido}

Here we consider that 
superconductivity in systems with the orbital degrees of freedom 
is affected primarily by 
splitting enegry between lowest and excited states, 
while kind of excited Kramers doublet may play secondary role 
to determine details of electronic properties. 
Based on this belief, in the following, 
we consider only $|\Gamma_{8} \rangle$ states. 
In other words, 
one $\Gamma_7$ state is assumed to be 
the highest energy state. 
Note that $|\Gamma_{8\pm}^{(1)} \rangle$ and
$|\Gamma_{8\pm}^{(2)} \rangle$ belong to $\Gamma_7$
and $\Gamma_6$ irreducible representations, respectively,
in the tetragonal system.
Although this assumption is not exactly the same as 
the level scheme obtained from 
experimental results mentioned above, 
using $\Gamma_6$ and $\Gamma_7$ states 
instead of two $\Gamma_7$ states 
will provide even reasonable result 
to discuss superconductivity of CeTIn$_5$. \cite{footnote1}
We stress that 
the Hamiltonian constructed from $\Gamma_8$ quartet is the simplest
possible model including essential physics of interplay between
pseudo-spin and orbital degrees of freedom.

We include itinerant features of $f$-electrons by 
considering nearest-neighbor hopping of $f$-electrons. 
Since realistic effective hopping of $f$-electrons 
includes that through 
hybridization with conduction electrons, 
such effects may be renormalized 
to the nearest-neighbor hopping of $f$-electrons 
after the conduction electron degrees of freedom are integrated out.
In the present case, the matrix elements of the hoppings 
depend on not only the hopping directions but also kinds of orbitals, 
since the forms of the wave functions of $|\Gamma_{8\pm}^{(1)} \rangle$
and $|\Gamma_{8\pm}^{(2)} \rangle$ states are different from each other. 
Noting that CeTIn$_5$ has a tetragonal crystal structure
and quasi two-dimensional Fermi surfaces
have been experimentally observed, \cite{Haga,Settai}
it is natural to consider the two-dimensional square lattice
composed of Ce$^{3+}$ ions.
Considering these points, 
we have estimated the hopping matrix elements 
through the $\sigma$-bond ($ff\sigma$) 
using the tight-binding method. \cite{Maehira,manybody2}
The matrix elements of 
nearest-neighbor hopping of $f$-electrons
$t^{\bf a}_{\tau\tau'}$ between
$\tau$ and $\tau'$ orbitals along the ${\bf a}$-direction
are given by
\begin{equation}
  t^{\bf x}_{11}=-\sqrt{3}t^{\bf x}_{12}=
  -\sqrt{3}t^{\bf x}_{21}=3t^{\bf x}_{22}=t,
\end{equation}
for ${\bf a}$=${\bf x}$, and
\begin{equation}
  t^{\bf y}_{11}=\sqrt{3}t^{\bf y}_{12}=
  \sqrt{3}t^{\bf y}_{21}=3t^{\bf y}_{22}=t, 
\end{equation}
for ${\bf a}$=${\bf y}$. \cite{Dagotto}
We use $t$=1 as energy unit in the following.

By further adding the on-site Coulomb interaction terms
among $f$-electrons, 
an effective Hamiltonian 
of CeTIn$_5$ compounds 
with orbital degrees of freedom 
is obtained as
\begin{eqnarray}
  H &=& \sum_{{\bf ia}\tau\tau'\sigma}
  t^{\bf a}_{\tau\tau'} f_{{\bf i}\tau\sigma}^{\dag}
                        f_{{\bf i+a}\tau'\sigma}
  -\Delta \sum_{\bf i}(n_{{\bf i}1\sigma}-n_{{\bf i}2\sigma})/2 
  \nonumber \\
  &+&U \sum_{{\bf i}\tau}n_{{\bf i}\tau\uparrow}
                         n_{{\bf i}\tau\downarrow}
  +U'\sum_{{\bf i}\sigma\sigma'}n_{{\bf i}1\sigma}
                                n_{{\bf i}2\sigma'},
\end{eqnarray}
where $f_{{\bf i}\tau\sigma}$ is the annihilation operator for
an $f$-electron with pseudo-spin $\sigma$ in the $\tau$-orbital state
$\Gamma_{8}^{(\tau)}$ at site ${\bf i}$,
${\bf a}$ is the vector connecting nearest-neighbor sites,
and $n_{{\bf i}\tau\sigma}$=$f_{{\bf i}\tau\sigma}^{\dag}
f_{{\bf i}\tau\sigma}$.
The first term represents the nearest-neighbor hopping of $f$-electrons.
The second term expresses the tetragonal CEF, represented by an
energy splitting $\Delta$ between the two orbitals.
In the third and fourth terms, $U$ and $U'$ are the intra- and
inter-orbital Coulomb interactions, respectively.
We ignore the Hund's rule coupling, since it may be irrelevant
for the quarter-filling case with one $f$-electron per site.
To keep the rotational invariance in the orbital space for
the interaction part of the Hamiltonian, 
$U'$ should be equal to $U$ when
we ignore the Hund's rule coupling. 
Thus, in this paper, we restrict ourselves to the case of $U$=$U'$.

Considering property of the hopping matrix element, \cite{Dagotto}
the present Hamiltonian in the quarter-filling 
may be regarded as an effective model in the hole-picture 
for the CuO$_2$ plane of 
a parent compound of high-$T_{\rm c}$ cuprate La$_2$CuO$_4$, 
although the practical value of $\Delta$ may be considerably large 
for the $d$-electron system. 
This fact means that some of the results obtained by using the present 
Hamiltonian can be used for the cuprate 
with a suitable choice of the parameter set. 
We also note that in the quarter-filling case,
the present model is reduced to a half-filled single-orbital
Hubbard model in the limit of $\Delta$=$\infty$.

Here we briefly discuss symmetry properties of the present Hamiltonian. 
Since all pseudo-spin operators 
$S^{\alpha}=(1/2)\sum_{{\bf i},\tau}\sum_{\sigma,\sigma'}
f_{{\bf i}\tau\sigma}^{\dag}\tau^{\alpha}_{\sigma\sigma'}
f_{{\bf i}\tau\sigma'}$ ($\alpha$=$x$, $y$, $z$) 
with $\hat{\tau}^{\alpha}$ being the Pauli matrices 
commute with this Hamiltonian, 
the system is invariant with respect to the rotation 
in the pseudo-spin space. 
Since the Hamiltonian is consisted of 
the pairs of creation and annihilation operators 
corresponding to each Kramers doublet, 
it is easily confirmed that the Hamiltonian is invariant 
under the time reversal. 
Since we consider the system of two-dimensional square lattice composed of 
Ce$^{3+}$-ion, the Hamiltonian commutes with all elements of 
the $D_{4h}$ point group. 
Obviously, the Hamiltonian is $U(1)$-gauge invariant. 
Thus, the Hamiltonian has the symmetry of 
$D_{4h} \times SU(2) \times \Theta \times U(1)$ after all 
where $SU(2)$ describes the rotation group in the 
pseudo-spin space and $\Theta$ the time reversal symmetry. 
In the following, we call pseudo-spin as ``spin" for simplicity.

\section{Formulation}

In our previous work, we have developed a weak-coupling theory for
superconductivity based on the same orbital degenerate model described above, 
using the static spin and orbital fluctuations obtained
within the random phase approximation (RPA). \cite{manybody}
On the other hand, 
considerably large discontinuity of the specific heat 
at the superconducting transition temperature 
has been observed in CeCoIn$_5$. \cite{CeCoIn5}
Since the discontinuity is much larger than the specific heat 
just above the superconducting transition temperature, 
the mass enhancement due to the strong interaction between $f$-electrons 
may not be responsible for the large discontinuity. 
Rather, this experimental fact seems to require 
the strong-coupling theory 
in order to understand the superconductivity in CeTIn$_5$ compounds. 
Thus, we should develop a strong-coupling theory for 
superconductivity based on the orbital degenerate model. 

In the present paper, we apply the fluctuation exchange (FLEX) 
approximation \cite{FLEX}
to the orbital degenerate model discussed in the preceding section. 
The FLEX approximation provides the Dyson-Gorkov equation 
where the normal and anomalous self-energies are obtained 
on an equal footing, namely a kind of the strong-coupling theory. 
Here, it should be noted that the strong-coupling theory 
includes two major changes for the orbital degenerate system 
compared with the single orbital case. 
One is multi-component nature of the superconducting order parameters 
in the orbital space, namely 
orbital symmetric and orbital antisymmetric gap functions. 
The other is the effect of mode-mode coupling 
among spin and orbital fluctuations. 
We expect that the mode-mode coupling significantly affects 
the temperature and frequency dependences of spin and orbital 
fluctuations. 
Therefore, 
in order to discuss superconductivity induced by these fluctuations 
the mode-mode coupling effect is very important. 
In the following, first
we discuss general relations satisfied by the Green's functions 
required from the symmetry of the system, 
then develop the scheme of the FLEX approximation 
for the degenerate model.

\subsection{Definition and Properties of Green's Function}
The normal Green's functions 
$G^{\sigma}_{mn}({\bf k},\tau)$ describing the propagating process 
of electrons from $n$-orbital to $m$-orbital 
with moment ${\bf k}$ and spin $\sigma$
and the anomalous ones $F^{\sigma\sigma'}_{mn}({\bf k},\tau)$ and 
$\overline{F}^{\sigma\sigma'}_{mn}({\bf k},\tau)$ 
describing the superconducting condensation 
are defined as
\begin{eqnarray}
   &&G^{\sigma}_{mn}({\bf k},\tau)=
   -\langle T_{\tau}[f_{{\bf k}m\sigma}(\tau)
                    f_{{\bf k}n\sigma}^{\dag}(0)] \rangle,\\
   &&F^{\sigma\sigma'}_{mn}({\bf k},\tau)=
   -\langle T_{\tau}[f_{{\bf k}m\sigma}(\tau)
                    f_{-{\bf k}n\sigma'}(0)] \rangle,\\
   &&\overline{F}^{\sigma\sigma'}_{mn}({\bf k},\tau)=
   -\langle T_{\tau}[f_{-{\bf k}m\sigma}^{\dag}(\tau)
                     f_{{\bf k}n\sigma'}^{\dag}(0)] \rangle,
\end{eqnarray}
where 
$f_{{\bf k}m\sigma}(\tau)=e^{(H-\mu N)\tau}f_{{\bf k}m\sigma}
e^{-(H-\mu N)\tau}$ with $\tau$ being the imaginary time, 
$N=\sum_{{\bf i},m,\sigma}f_{{\bf i}m\sigma}^{\dag}
                        f_{{\bf i}m\sigma}$ 
is an operator of the $f$-electron number, 
$\mu$ the chemical potential, 
and $T_{\tau}$ describes the time ordered product. 
In these equations, 
$\langle\cdots\rangle$ means the thermodynamical average. 
It is convenient to transform the imaginary time Green's functions 
to the frequency representation given by
\begin{equation}
   {\cal G}({\bf k},{\rm i}\omega_{l})
   =\int^{1/T}_{0}d\tau\hspace{1mm}e^{{\rm i}\omega_{l}\tau}
    {\cal G}({\bf k},\tau),
\end{equation}
where ${\cal G}({\bf k},\tau)$ represents a component of the normal or 
anomalous Green's functions defined above, and 
$\omega_{l}$=$(2l+1)\pi T$ is 
the Matsubara frequency for fermions. 

Then, 
under the assumption that the superconducting transition 
does not break the time reversal symmetry 
for the orbital degenerate system, 
Green's functions satisfy the relations as
\begin{eqnarray}
  &&G^{\sigma}_{mn}({\bf k},{\rm i}\omega_{l})
  =G^{\sigma}_{nm}({\bf k},-{\rm i}\omega_{l})^{\ast}
  =G^{\overline{\sigma}}_{nm}(-{\bf k},{\rm i}\omega_{l}),\\
  &&F^{\sigma\sigma'}_{mn}({\bf k},{\rm i}\omega_{l})
  =-F^{\sigma'\sigma}_{nm}(-{\bf k},-{\rm i}\omega_{l})\nonumber\\
  &=&\sigma\sigma'F^{\overline{\sigma}\overline{\sigma'}}_{mn}
       (-{\bf k},-{\rm i}\omega_{l})^{\ast}
  =\sigma\sigma'\overline{F}^{\overline{\sigma'}\overline{\sigma}}_{nm}
       (-{\bf k},{\rm i}\omega_{l}), 
\end{eqnarray}
where the last equalities in these relations
are obtained by the time reversal invariance. 
Due to the $SU(2)$-symmetry in the spin space, 
$F^{\sigma\sigma'}_{mn}({\bf k},{\rm i}\omega_{l})$
are decomposed into the spin-singlet and spin-triplet 
irreducible representations, defined as
\begin{equation}
  F^{\rm s}_{mn}({\bf k},{\rm i}\omega_{l})
  =\frac{1}{2}
  (F^{\uparrow\downarrow}_{mn}({\bf k},{\rm i}\omega_{l})
  -F^{\downarrow\uparrow}_{mn}({\bf k},{\rm i}\omega_{l})),
\end{equation}
and
\begin{equation}
  F^{\rm t}_{mn}({\bf k},{\rm i}\omega_{l})
  =\frac{1}{2}
  (F^{\uparrow\downarrow}_{mn}({\bf k},{\rm i}\omega_{l})
  +F^{\downarrow\uparrow}_{mn}({\bf k},{\rm i}\omega_{l})),
\end{equation}
where $F^{\rm t}_{mn}({\bf k},{\rm i}\omega_{l})$ defined above 
is a representative of the three components of the spin-triplet pairs. 
In the paramagnetic system, 
since we can suppress a superscript describing spin state 
of $G^{\sigma}_{mn}({\bf k},{\rm i}\omega_{l})$, 
we obtain the relations 
for the normal Green's functions as
\begin{equation}
  G_{mn}({\bf k},{\rm i}\omega_{l})
  =G_{nm}({\bf k},-{\rm i}\omega_{l})^{\ast}
  =G_{nm}({\bf k},{\rm i}\omega_{l}).
\end{equation}
Furthermore, 
{\it {when the orbitals are defined at the Bravais lattice}}, 
the inversion operation changes only the wave vector $\bf k$ to -$\bf k$ 
independent on the orbital state as well as spin 
of the quasi-particles forming the Cooper pair. 
Thus, we obtain the relations 
for $F^{\xi}_{mn}({\bf k},{\rm i}\omega_{l})$ ($\xi$=s or t) as
\begin{equation}
  F^{\xi}_{mn}({\bf k},{\rm i}\omega_{l})
  =F^{\xi}_{mn}({\bf k},-{\rm i}\omega_{l})^{\ast}
  =F^{\xi}_{nm}({\bf k},{\rm i}\omega_{l})^{\ast}.
\end{equation}
Among 
$F^{\uparrow\downarrow}_{mn}({\bf k},{\rm i}\omega_{l})$, 
$\overline{F}^{\downarrow\uparrow}_{mn}({\bf k},{\rm i}\omega_{l})$, 
and $F^{\xi}_{mn}({\bf k},{\rm i}\omega_{l})$, 
we also obtain
\begin{equation}
  F^{\xi}_{mn}({\bf k},{\rm i}\omega_{l})
  =F^{\uparrow\downarrow}_{mn}({\bf k},{\rm i}\omega_{l})
  =\overline{F}^{\downarrow\uparrow}_{nm}({\bf k},{\rm i}\omega_{l}).
\end{equation}
Note that these relations are obtained 
in the time reversal invariant system 
regardless of the spin-singlet and 
even-parity pair or the spin-triplet and odd-parity pair. 
We emphasize that these relations are quite useful 
to make the formulation simple. 

By transforming 
$F^{\xi}_{mn}({\bf k},{\rm i}\omega_{l})$ ($\xi$=s or t) 
to the ``orbital-symmetric" and ``orbital-antisymmetric" representations, 
four real anomalous Green's functions are defined as
\begin{eqnarray}
  &&F^{\xi}_{1}({\bf k},{\rm i}\omega_{l})\equiv 
    F^{\xi}_{11}({\bf k},{\rm i}\omega_{l}), \\
  &&F^{\xi}_{2}({\bf k},{\rm i}\omega_{l})\equiv 
    F^{\xi}_{22}({\bf k},{\rm i}\omega_{l}), \\
  &&F^{\xi}_{3}({\bf k},{\rm i}\omega_{l})\equiv\frac{1}{\sqrt{2}}
    (F^{\xi}_{12}({\bf k},{\rm i}\omega_{l})
    +F^{\xi}_{21}({\bf k},{\rm i}\omega_{l})), \\
  &&F^{\xi}_{4}({\bf k},{\rm i}\omega_{l})\equiv\frac{\rm i}{\sqrt{2}}
    (F^{\xi}_{12}({\bf k},{\rm i}\omega_{l})
    -F^{\xi}_{21}({\bf k},{\rm i}\omega_{l})),
\end{eqnarray}
where $F^{\xi}_{4}({\bf k},{\rm i}\omega_{l})$ just describes 
the orbital-antisymmetric pairing. 
For the spin-singlet state, they satisfy the relations as
\begin{eqnarray}
  &&F^{\rm s}_{m}({\bf k},{\rm i}\omega_{l})
  =F^{\rm s}_{m}(-{\bf k},{\rm i}\omega_{l})
  =F^{\rm s}_{m}({\bf k},-{\rm i}\omega_{l}),\\
  &&F^{\rm s}_{4}({\bf k},{\rm i}\omega_{l})
  =F^{\rm s}_{4}(-{\bf k},{\rm i}\omega_{l})
  =-F^{\rm s}_{4}({\bf k},-{\rm i}\omega_{l}),
\end{eqnarray}
and for the spin-triplet state, we obtain
\begin{eqnarray}
  &&F^{\rm t}_{m}({\bf k},{\rm i}\omega_{l})
  =-F^{\rm t}_{m}(-{\bf k},{\rm i}\omega_{l})
  =F^{\rm t}_{m}({\bf k},-{\rm i}\omega_{l}),\\
  &&F^{\rm t}_{4}({\bf k},{\rm i}\omega_{l})
  =-F^{\rm t}_{4}(-{\bf k},{\rm i}\omega_{l})
  =-F^{\rm t}_{4}({\bf k},-{\rm i}\omega_{l}),
\end{eqnarray}
where $m$ represents the number of the orbital-symmetric component, 
namely $m$=1, 2, or 3. 
One can see that the orbital-antisymmetric anomalous Green's function 
has odd-frequency dependence, irrespective of 
the spin state of the Cooper-pair. 
Thus, due to the odd-frequency property, 
the orbital-antisymmetric component of the Cooper pair 
may not provide essential contribution to superconductivity. 
In addition to this feature in the frequency space, we should also 
pay attention to the $\bf k$-dependences of 
$F^{\xi}_{mn}({\bf k},{\rm i}\omega_{l})$ 
because the lattice system and the local wave functions 
should be rotated simultaneously. 
For example, a symmetry operation $C_4$ of the tetragonal point group 
rotates the wave vector ${\bf k}$, and also 
changes the sign of only the $\Gamma_{7}$ wave function of $f$-electron. 
The latter effect of the symmetry operation 
leads to the difference between $\bf k$-dependences of 
the orbital-diagonal 
and the orbital-offdiagonal anomalous Green's functions,  
in order to preserve the symmetry of superconductivity. 
After all, when the superconducting state belongs to $\Gamma$ 
irreducible representation, 
the $\bf k$-dependences of 
$F^{\xi}_{12}({\bf k},{\rm i}\omega_{l})$ and 
$F^{\xi}_{21}({\bf k},{\rm i}\omega_{l})$ have 
the $\Gamma\times B_{\rm 1g}$-symmetry
while the symmetry properties of 
$F^{\xi}_{11}({\bf k},{\rm i}\omega_{l})$ and 
$F^{\xi}_{22}({\bf k},{\rm i}\omega_{l})$ in the $\bf k$-space 
behave as the $\Gamma$ irreducible represenation, 
the same as the symmetry of the order parameter. 
These relations obtained in this subsection 
are useful for practical calculations.

Here we comment 
on difference between the strong-
and weak-coupling theories 
for the degenerate model. 
As will be obtained below, 
the superconducting gap functions 
are not independent for the orbital-symmetric 
and orbital-antisymmetric parts generally. 
On the other hand, 
in the weak-coupling theory of the same degenerate model, 
the orbital-symmetric components and the orbital-antisymmetric 
one of $\Sigma^{\xi(2)}_{mn}(k)$ define 
separate superconducting states from each other. \cite{manybody}
Considering the result in the present subsection, 
the complete separation within the weak-coupling theory 
is due to ignoring the odd-frequency dependence of 
the orbital-antisymmetric component of the anomalous Green's functions 
$F^{\xi}_{4}({\bf k},{\rm i}\omega_{l})$.

\subsection{FLEX Approximation for the Degenerate Model}
In order to develop a strong-coupling theory of 
superconductivity in the degenerate system, 
it is convenient to introduce the Dyson-Gorkov equations, 
which is in the matrix form because of the orbital degree of freedom, 
given by
\begin{eqnarray}
   &&\hat{G}(k)=\hat{G}^{(0)}(k)
   +\hat{G}^{(0)}(k)\hspace{1mm}\hat{\Sigma}^{(1)}(k)
    \hspace{1mm}\hat{G}(k)\nonumber\\
   &&\hspace*{24.5mm}+\hat{G}^{(0)}(k)\hspace{1mm}
    \hat{\Sigma}^{\xi(2)}(k)\hspace{1mm}
    \hat{F}^{\xi}(k)^{\rm t},\\
   &&\hat{F}^{\xi}(k)=
    \hat{G}^{(0)}(k)\hspace{1mm}\hat{\Sigma}^{(1)}(k)
    \hspace{1mm}\hat{F}^{\xi}(k)\nonumber\\
   &&\hspace*{11.5mm}-\hat{G}^{(0)}(k)\hspace{1mm}
    \hat{\Sigma}^{\xi(2)}(k)\hspace{1mm}
    \hat{G}(-k)^{\rm t},
\end{eqnarray}
where $\hat{\Sigma}^{(1)}(k)$ is a matrix of the normal self-energies, 
$\hat{\Sigma}^{\xi(2)}(k)$ describes 
a matrix of the anomalous self-energies for the spin-$\xi$ state pairing 
and the abbreviation $k\equiv({\bf k},{\rm i}\omega_{l})$ is used.
$\hat{G}^{(0)}(k)$ is the matrix of 
the noninteracting Green's function 
whose matrix elements are given by 
\begin{equation}
  G^{(0)}_{mn}({\bf k},{\rm i}\omega_{l})
  =\sum_{p}\alpha_{{\bf k}pm}\alpha_{{\bf k}pn}
   \frac{1}{{\rm i}\omega_{l}-E_{{\bf k}p}}
\end{equation}
with
\begin{eqnarray}
  &&E_{{\bf k}p}=\frac{\epsilon_{{\bf k}11}+\epsilon_{{\bf k}22}}{2}
  +(-1)^{p}
    \sqrt{\frac{(\epsilon_{{\bf k}11}-\epsilon_{{\bf k}22})^{2}}{4}
    +\epsilon_{{\bf k}12}^{2}},\\
  &&\alpha_{{\bf k}11}=\alpha_{{\bf k}22}
  =\left[1+\frac{E_{{\bf k}2}-\epsilon_{{\bf k}11}}
            {E_{{\bf k}2}-\epsilon_{{\bf k}22}}\right]^{-1/2},\\
  &&\alpha_{{\bf k}12}=-\alpha_{{\bf k}21}
  =\frac{\epsilon_{{\bf k}12}}{E_{{\bf k}2}-\epsilon_{{\bf k}22}}
   \alpha_{{\bf k}11},
\end{eqnarray}
and
\begin{eqnarray}
  &&\epsilon_{{\bf k}11}=2\hspace{1mm}t\hspace{1mm}(\cos{k_{x}}+\cos{k_{y}})
    -\frac{\Delta}{2}-\mu,\\
  &&\epsilon_{{\bf k}22}=\frac{2}{3}\hspace{1mm}t\hspace{1mm}
    (\cos{k_{x}}+\cos{k_{y}})+\frac{\Delta}{2}-\mu,\\
  &&\epsilon_{{\bf k}12}=\epsilon_{{\bf k}21}
    =-\frac{2}{\sqrt{3}}\hspace{1mm}t\hspace{1mm}(\cos{k_{x}}-\cos{k_{y}}),
\end{eqnarray}
where $E_{{\bf k}p}$ is the energy dispersion of $p$-th band 
and $\alpha_{{\bf k}pm}$ is the weight of $m$-th orbital 
for the $p$-th band at wave vector ${\bf k}$. 

In order to obtain a concrete form of the matrix elements 
of the self-energy for the degenerate model, 
we adopt the FLEX approximation. 
First, we define a ``Luttinger-Ward" functional $\Phi[{\cal G}(k)]$ 
consisting of ladder and bubble diagrams for particle-hole processes 
but ignoring particle-particle processes, 
and then generate matrix elements 
of the self-energy by functional differentiation of $\Phi[{\cal G}(k)]$ 
with respect to ${\cal G}(k)$ where ${\cal G}(k)$ represents a component 
of the normal and/or anomalous Green's functions. 
By this procedure, the self-energies are obtained 
within the FLEX approximation by
\begin{equation}
   \Sigma^{(1)}_{mn}(k)=\frac{T}{N_0}\sum_{q}\sum_{\mu\nu}
    T^{\rm eff}_{\mu m,\nu n}(q)G_{\mu\nu}(k-q),
\end{equation}
and
\begin{equation}
   \Sigma^{\xi(2)}_{mn}(k)=\frac{T}{N_0}\sum_{q}\sum_{\mu\nu}
    T^{\xi}_{\mu m,n\nu}(q)F^{\xi}_{\mu\nu}(k-q),
\end{equation}
where an abbreviation $q\equiv({\bf q},{\rm i}\Omega_{l})$ is used 
with the boson Matsubara frequency $\Omega_{l}$=$2l\pi T$.
The fluctuation exchange interactions 
used in $\Sigma^{(1)}_{mn}(k)$ are given by
\begin{eqnarray}
   &&T^{\rm eff}_{\mu m,\nu n}(q)
   =\frac{1}{2}[3\hat{U}^{\rm s}\hat{\chi}^{\rm s}(q)\hat{U}^{\rm s}
    +\hat{U}^{\rm o}\hat{\chi}^{\rm o}(q)\hat{U}^{\rm o}\nonumber\\
   &&\hspace{2mm}-\frac{1}{2}(\hat{U}^{\rm s}+\hat{U}^{\rm o})
     \hat{\overline{\chi}}^{\sigma\sigma}(q)(\hat{U}^{\rm s}+\hat{U}^{\rm o})
    +3\hat{U}^{\rm s}-\hat{U}^{\rm o}]_{\mu m,\nu n}.
\end{eqnarray}
For the spin-singlet channel, 
matrix elements of the effective pairing interaction are given by
\begin{eqnarray}
   &&T^{\rm s}_{\mu m,n\nu}(q)
   =\frac{1}{2}[3\hat{U}^{\rm s}\hat{\chi}^{\rm s}(q)\hat{U}^{\rm s}
    -\hat{U}^{\rm o}\hat{\chi}^{\rm o}(q)\hat{U}^{\rm o}\nonumber\\
   &&\hspace{2mm}+\frac{1}{2}(\hat{U}^{\rm s}+\hat{U}^{\rm o})
     \hat{\overline{\chi}}^{\sigma\overline{\sigma}}(q)
    (\hat{U}^{\rm s}+\hat{U}^{\rm o})
    +\hat{U}^{\rm s}+\hat{U}^{\rm o}]_{\mu m,n\nu},
\end{eqnarray}
and for the spin-triplet channel, we obtain
\begin{eqnarray}
   &&T^{\rm t}_{\mu m,n\nu}(q)
   =\frac{1}{2}[-\hat{U}^{\rm s}\hat{\chi}^{\rm s}(q)\hat{U}^{\rm s}
    -\hat{U}^{\rm o}\hat{\chi}^{\rm o}(q)\hat{U}^{\rm o}\nonumber\\
   &&\hspace{2mm}+\frac{1}{2}(\hat{U}^{\rm s}+\hat{U}^{\rm o})
     \hat{\overline{\chi}}^{\sigma\overline{\sigma}}(q)
    (\hat{U}^{\rm s}+\hat{U}^{\rm o})
    +\hat{U}^{\rm s}+\hat{U}^{\rm o}]_{\mu m,n\nu},
\end{eqnarray}
with
\begin{eqnarray}
  \hat{U}^{\rm s}
 =\left[\begin{array}{cccc}
   U & 0 & 0 & 0 \\
   0 & U & 0 & 0 \\
   0 & 0 & U' & 0\\
   0 & 0 & 0 & U'
        \end{array}\right],\hspace{1mm}
  \hat{U}^{\rm o}
 =\left[\begin{array}{cccc}
   U & 2U' & 0 & 0 \\
   2U' & U & 0 & 0 \\
   0 & 0 & -U' & 0\\
   0 & 0 & 0 & -U'
        \end{array}\right].\nonumber
\end{eqnarray}
As has alredy been pointed out in the previous work, \cite{manybody}
one can see that the contributions 
to the pairing interaction of the spin and orbital fluctuations 
are, in general, destructive 
for the spin-singlet channel while constructive 
for the spin-triplet channel. 
$\hat{\chi}^{\rm s}(q)$ and $\hat{\chi}^{\rm o}(q)$
in the above expressions of self-energies correspond to 
the spin and orbital fluctuations, respectively, 
and within the FLEX approximation these are given as
\begin{eqnarray}
  &&\hat{\chi}^{\rm s}(q)=
  [\hat{1}-\hat{U}^{\rm s}\hat{\overline{\chi}}^{\rm s}(q)]^{-1}
  \hat{\overline{\chi}}^{\rm s}(q),\\
  &&\hat{\chi}^{\rm o}(q)=
  [\hat{1}+\hat{U}^{\rm o}\hat{\overline{\chi}}^{\rm o}(q)]^{-1}
  \hat{\overline{\chi}}^{\rm o}(q),
\end{eqnarray}
with
\begin{eqnarray}
  &&\hat{\overline{\chi}}^{\rm s}(q)=
    \hat{\overline{\chi}}^{\sigma\sigma}(q)
   -\hat{\overline{\chi}}^{\sigma\overline{\sigma}}(q),\\
  &&\hat{\overline{\chi}}^{\rm o}(q)=
    \hat{\overline{\chi}}^{\sigma\sigma}(q)
   +\hat{\overline{\chi}}^{\sigma\overline{\sigma}}(q),
\end{eqnarray}
where $\hat{\overline{\chi}}^{\sigma\sigma'}(q)$ is 
the matrix of the irreducible susceptibility corresponding to 
one bubble Feynman diagram with using the renormalized 
Green's functions, given by
\begin{eqnarray}
  &&\hat{\overline{\chi}}^{\sigma\sigma'}(q)\nonumber\\
 &=&\left[\begin{array}{cccc}
   \overline{\chi}^{\sigma\sigma'}_{1111}(q) & 
   \overline{\chi}^{\sigma\sigma'}_{1122}(q) & 
   \overline{\chi}^{\sigma\sigma'}_{1112}(q) & 
   \overline{\chi}^{\sigma\sigma'}_{1121}(q) \\
   \overline{\chi}^{\sigma\sigma'}_{2211}(q) & 
   \overline{\chi}^{\sigma\sigma'}_{2222}(q) & 
   \overline{\chi}^{\sigma\sigma'}_{2212}(q) & 
   \overline{\chi}^{\sigma\sigma'}_{2221}(q) \\
   \overline{\chi}^{\sigma\sigma'}_{1211}(q) & 
   \overline{\chi}^{\sigma\sigma'}_{1222}(q) & 
   \overline{\chi}^{\sigma\sigma'}_{1212}(q) & 
   \overline{\chi}^{\sigma\sigma'}_{1221}(q) \\
   \overline{\chi}^{\sigma\sigma'}_{2111}(q) & 
   \overline{\chi}^{\sigma\sigma'}_{2122}(q) & 
   \overline{\chi}^{\sigma\sigma'}_{2112}(q) & 
   \overline{\chi}^{\sigma\sigma'}_{2121}(q)
        \end{array}\right].
\end{eqnarray}
Here we define
\begin{equation}
  \overline{\chi}^{\sigma\sigma}_{ij,st}(q)
   =-\frac{T}{N_0}\sum_{k}G_{si}(k+q)G_{jt}(k),
\end{equation}
and
\begin{equation}
  \overline{\chi}^{\sigma\overline{\sigma}}_{ij,st}(q)
   =\frac{T}{N_0}\sum_{k}F^{\xi}_{ti}(k+q)F^{\xi}_{js}(k).
\end{equation}
Although these expressions are derived 
for the doubly degenerate system, 
it is straightforward to extend this formalism to systems 
with more orbital degrees of freedom. 

To calculate $T_{\rm c}$ we linearize 
the Dyson-Gorkov equations with respect to 
$\hat{F}^{\xi}(k)$ or $\hat{\Sigma}^{\xi(2)}(k)$. 
The transition temperature for superconductivity is determined 
as the temperature below which the linearized equation for 
$\hat{\Sigma}^{\xi(2)}(k)$ has a nontrivial solution. 
The linearized gap equation is given by 
\begin{equation}
   \Sigma^{\xi(2)}_{mn}(k)
   =-\frac{T}{N_0}\sum_{q}\sum_{s t}
    W^{\xi}_{mn,st}(q)\Sigma^{\xi(2)}_{s t}(k-q)
\end{equation}
with
\begin{equation}
   W^{\xi}_{mn,st}(q)=
    \sum_{\mu\nu}T^{\xi}_{\mu m,n\nu}(q)G_{\mu s}(k-q)G_{\nu t}(q-k).
\end{equation}



Finally, we mention actual calculations in the FLEX approximation. 
The FLEX calculation is numerically carried out 
at fixed parameter values of $U$=$U'$=4 and 
one $f$-electron per site on the average.
Summations involved in the above self-consistent equations are
performed using the fast Fourier transformation algorithm both 
for the ${\bf k}$-space with $32\times32$ meshes 
in the first Brillouin zone 
and for Matsubara frequency sum. 
In particular, with respect to the Matsubara frequency sum, 
we have adopted an useful method developed 
by Deisz $et$ $al$. \cite{Deisz} 
to include high frequency contribution efficiently. 
When the relative error of every matrix element of 
$\hat{\Sigma}^{(1)}(k)$ for all ${\bf k}$ and $\omega_{l}$ becomes 
smaller than 10$^{-6}$, we assume that the solution is obtained 
for the self-consistent equations. 
In the present calculations, 
a second-order magnetic transition is defined by
\begin{equation}
  {\rm det}
  [\hat{1}-\hat{U}^{\rm s}\hat{\overline{\chi}}^{\rm s}(q)]
  =\eta,
\end{equation}
where $\eta=0.002$ is used throughout this paper. 
Introduction of such a small value for $\eta$ 
may be understood as the effect of weak three-dimensionality 
which is ignored in the present treatment. 
In general, increasing three-dimensionality 
extends the antiferromagnetic phase, 
while it suppresses the superconducting phase. \cite{Takimoto2}
A choice of sufficiently small value of $\eta$ is considered 
to be consistent 
with a description of the quasi-two dimensional system 
such as CeTIn$_5$ compounds.


\section{Calculated results}
\subsection{Spin and Orbital Fluctuations}
In this section, we show the results obtained 
within the FLEX approximation 
described in the previous section. 
We start from properties of the spin and orbital fluctuations 
of the orbital degenerate model. 
Here, in order to characterize the strength of the spin and orbital 
fluctuations, we define $\alpha^{\rm s}$ and $\alpha^{\rm o}$, 
respectively, as
\begin{eqnarray}
  &&\alpha^{\rm s}={\rm Min}\hspace{1mm}{\rm det}
  [\hat{1}-\hat{U}^{\rm s}
   \hat{\overline{\chi}}^{\rm s}({\bf q},0)],\\
  &&\alpha^{\rm o}={\rm Min}\hspace{1mm}{\rm det}
  [\hat{1}+\hat{U}^{\rm o}
   \hat{\overline{\chi}}^{\rm o}({\bf q},0)],
\end{eqnarray}
where ``Min" in these expressions means the minimum value 
in the $\bf q$-space. 
These quantities indicate inverses of enhancement factors 
of dominant spin and orbital fluctuations in the momentum space. 
Note that decreases of $\alpha^{\rm s}$ and $\alpha^{\rm o}$ correspond 
to developments of the spin and orbital fluctuations, respectively, 
at a wave vector in the ${\bf q}$-space. 
In the present approximation, vanishing of an eigenvalue of 
$[\hat{1}-\hat{U}^{\rm s}
   \hat{\overline{\chi}}^{\rm s}({\bf q},0)]$ or 
$[\hat{1}+\hat{U}^{\rm o}
   \hat{\overline{\chi}}^{\rm o}({\bf q},0)]$ 
defines an instability of 
the spin or orbital density fluctuation. 
Thus, $\alpha^{\rm s}$ and $\alpha^{\rm o}$ 
may be used as indicators of the strength 
of spin and orbital fluctuations. 
The temperature dependence of $\alpha^{\rm s}$ and $\alpha^{\rm o}$ 
are shown in Fig.~1 for various values of the orbital splitting 
energy $\Delta$. 
With decreasing the temperature, $\alpha^{\rm s}$ decreases 
while $\alpha^{\rm o}$ is almost independent of the temperature. 
By increasing the orbital splitting energy $\Delta$, 
$\alpha^{\rm s}$ is considerably suppressed 
at sufficiently low temperatures 
while $\alpha^{\rm o}$ increases.

\begin{figure}[t]
\centerline{\epsfxsize=6.0truecm \epsfbox{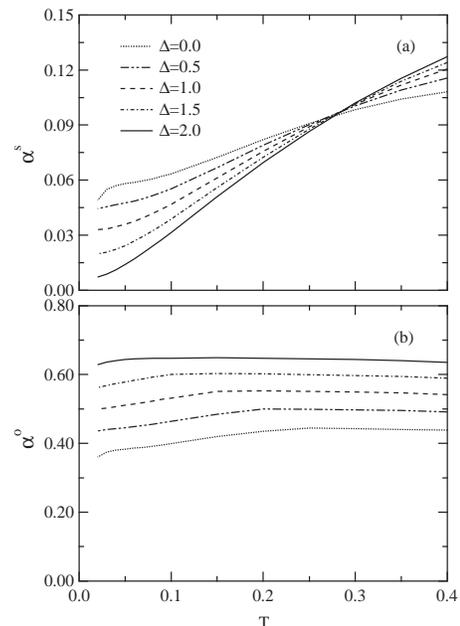} }
\caption{Temperature dependences of (a) $\alpha^{\rm s}$ 
and (b) $\alpha^{\rm o}$ for each value of $\Delta$.}
\end{figure}

In order to investigate the dynamical properties of 
the spin and orbital fluctuations, we define 
local spin and orbital susceptibilities as
\begin{equation}
  {\rm Im}\chi^{\rm s(o)}_{\mu\nu,mn}(\omega)=\frac{1}{N_0}\sum_{\bf q}
   {\rm Im}\chi^{\rm s(o)}_{\mu\nu,mn}({\bf q}, \omega+{\rm i}\delta),
\end{equation}
where $\delta$ is a positive infinitesimal quantity. 
The analytical continuation of $\chi^{\rm s(o)}_{\mu\nu,mn}(q)$ 
to the real axis is carried out by the Pad$\acute{\rm e}$ approximants. 
In order to clarify contributions to the longitudinal and transverse 
components of the orbital fluctuation 
from ${\rm Im}\chi^{\rm o}_{\mu\nu,mn}(\omega)$, 
we introduce operators of the charge density, 
longitudinal orbital density, and transverse orbital density as 
$\rho_{\bf i}=(1/2)\sum_{m,\sigma}f_{{\bf i}m\sigma}^{\dag}
                        f_{{\bf i}m\sigma}$, 
$\tau_{\bf i}^{L}=(1/2)\sum_{m,n,\sigma}f_{{\bf i}m\sigma}^{\dag}
                  \tau^{z}_{mn}f_{{\bf i}n\sigma}$, and 
$\tau_{\bf i}^{T}=\sum_{m,n,\sigma}f_{{\bf i}m\sigma}^{\dag}
    (\hat{\tau}^{x}+{\rm i}\hat{\tau}^{y})_{mn}f_{{\bf i}n\sigma}$, 
respectively. 
Dynamical susceptibilities for these operators are defined as
\begin{eqnarray}
  &&\chi^{\rm o}_{\rm C}(\omega)
    ={\rm i}\int^{\infty}_{0}dt\hspace{1mm}
     e^{{\rm i}\omega t-\delta t}
    \langle [\rho_{\bf i}(t), \rho_{\bf i}(0)] \rangle,\\
  &&\chi^{\rm o}_{\rm L}(\omega)
    ={\rm i}\int^{\infty}_{0}dt\hspace{1mm}
     e^{{\rm i}\omega t-\delta t}
    \langle [\tau_{\bf i}^{L}(t), \tau_{\bf i}^{L}(0)] \rangle,\\
  &&\chi^{\rm o}_{\rm T}(\omega)
    ={\rm i}\int^{\infty}_{0}dt\hspace{1mm}
     e^{{\rm i}\omega t-\delta t}
    \langle [\tau_{\bf i}^{T}(t), \tau_{\bf i}^{T}(0)] \rangle,
\end{eqnarray}
where these quantities correspond to the local components of 
the net charge fluctuation, the longitudinal orbital fluctuation, 
and the transverse orbital fluctuation, respectively. 
By using $\chi^{\rm o}_{\mu\nu,mn}(\omega)$,
these dynamical susceptibilities are described as
\begin{eqnarray}
  &&\chi^{\rm o}_{\rm C}(\omega)
  =\frac{1}{2}
   [\chi^{\rm o}_{11,11}(\omega)+\chi^{\rm o}_{22,22}(\omega)
   +2\chi^{\rm o}_{11,22}(\omega)],\\
  &&\chi^{\rm o}_{\rm L}(\omega)
  =\frac{1}{2}
   [\chi^{\rm o}_{11,11}(\omega)+\chi^{\rm o}_{22,22}(\omega)
   -2\chi^{\rm o}_{11,22}(\omega)],\\
  &&\chi^{\rm o}_{\rm T}(\omega)
  =\chi^{\rm o}_{21,21}(\omega),
\end{eqnarray}
where 
we make use of relation 
$\chi^{\rm o}_{11,22}(\omega)=\chi^{\rm o}_{22,11}(\omega)$. 
The frequency dependence of 
${\rm Im}\chi^{\rm s}_{\mu\nu,mn}(\omega)$ 
(${\rm Im}\chi^{\rm o}_{\rm X}(\omega)$)
for $\Delta/t=0$ and $\Delta/t=2$
are shown in Fig.~2(a) and Fig.~2(b) 
(in Fig.~2(c) and Fig.~2(d)), respectively. 
From these figures, one can see that 
the intensity of ${\rm Im}\chi^{\rm s}_{11,11}(\omega)$ 
in low-energy region develops with increasing $\Delta$ 
while intensities of two components of the orbital fluctuations 
are suppressed and net charge fluctuation almost unchanged. 
In the present case, it is understood that 
the spin fluctuation of the $f$-electron belonging to $\tau$=1 orbital 
(corresponding to $\Gamma_{8}^{(1)}$) provides predominant 
contribution compared with other fluctuations. 
The longitudinal orbital fluctuation seems to play 
the secondary role.

Here we comment on the reason why 
intensities of the orbital fluctuations in the low-energy region 
do not develop significantly even for the case of $\Delta$=0. 
We note that any spectral functions 
satisfy the sum rule in the frequency space. 
It means that 
suppression of the spectrum in the low-energy region 
leads to enhancement of that in the high-energy region 
or vice versa. 
In the present Hamiltonian, 
the $f$-electron number in the 
$\tau$=1 orbital $n_1$ is larger than $n_2$ 
due to the difference of the strength of hopping integrals 
included in the kinetic term 
$\epsilon_{{\bf k}11}$ and $\epsilon_{{\bf k}22}$. 
Since $n_{1}-n_{2}\ne 0$ in this case, 
the orbital polarization represented by $\tau_{\bf i}^{L}$ 
is non-zero, 
the longitudinal orbital fluctuation should be suppressed 
by the factor of $\langle\tau_{\bf i}^{L}\rangle^{2}$. 
This is the reason why the spectrum of the longitudinal 
orbital fluctuation in the low-energy region 
is suppressed even for the case of $\Delta$=0. 

In order to discuss the nature of transverse orbital fluctuation, 
we derive the asymptotic forms of 
$\chi^{\rm s(o)}_{\mu\nu,mn}({\bf q}, z)$ for $z\gg 1 $ as
\begin{equation}
  \chi^{\rm s(o)}_{\mu\nu,mn}({\bf q}, z)\approx
    -\delta\chi^{\rm s(o)}_{\mu\nu,mn}({\bf q})\frac{1}{z}
    +O(\frac{1}{z^2}),
\end{equation}
with coefficients 
\begin{equation}
  \delta\chi^{\rm s(o)}_{12,12}({\bf q})
    =-\delta\chi^{\rm s(o)}_{21,21}({\bf q})
    =\frac{1}{2}(n_{1}-n_{2}),
\end{equation}
for ($\mu\nu,mn$)=(12,12) or (21,21) 
and
\begin{equation}
  \delta\chi^{\rm s(o)}_{\mu\nu,mn}({\bf q})=0
\end{equation}
for other cases. 
Note that $\chi^{\rm o}_{21,21}({\bf q}, z)$ is just equal to 
$\chi^{\rm o}_{\rm T}({\bf q}, z)$. 
From these expressions, we can conclude that 
in the present case of $n_{1}-n_{2}\ne 0$, 
the intensity of spectrum for the transverse orbital fluctuation 
shifts to higher energy region to satisfy the sum rule. 
Thus, the significant difference of the $f$-electron numbers 
for the two orbitals 
prevents essential development of the low-energy excitation 
for the orbital fluctuations. 
By noting that the value of $n_{1}-n_{2}$ is controlled by 
the orbital splitting energy $\Delta$ included in 
$\epsilon_{{\bf k}11}$ and $\epsilon_{{\bf k}22}$, 
it is considered that not only the longitudinal orbital fluctuation 
but also the transverse orbital fluctuation are suppressed by 
increasing the orbital splitting energy.

\begin{figure}[t]
\centerline{\epsfxsize=8.5truecm \epsfbox{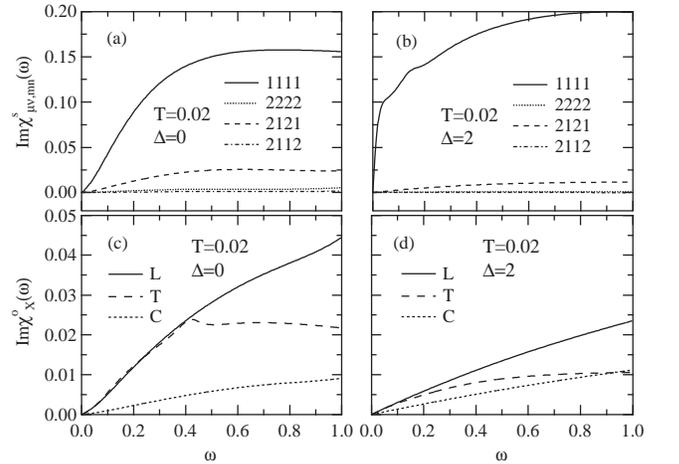} }
\caption{Spectra of the local components of 
(a) spin  and (c) orbital susceptibilities for $\Delta$=0. 
(b) and (d) are for $\Delta$=2. 
For (a) and (b), solid, dotted, dashed, and dash-dotted lines 
describe components of $\mu\nu mn$=1111, 2222, 2121, and 2112, 
respectively. 
For (c) and (d), solid, dashed, and dotted lines 
correspond to X=L, T, and C, 
respectively. }
\end{figure}

The momentum, ${\bf q}$, dependences of the principal components of 
$\hat{\chi}^{\rm s}({\bf q},0)$ and $\hat{\chi}^{\rm o}({\bf q},0)$ 
are shown in Fig.~3, at a fixed temperature 
$T$=0.02 for different orbital splitting energy, $\Delta$. 
The upper and lower panels indicate the results 
for $\Delta$=0 and $\Delta$=2, respectively. 
For $\Delta$=0, 
the antiferromagnetic spin fluctuation in the $\tau$=1 orbital 
corresponding to $\Gamma_{8}^{(1)}$ 
is enhanced, but not sufficiently developed to induce 
$d_{x^2-y^2}$-wave superconductivity.
With increasing the orbital splitting energy to $\Delta$=2,
the antiferromagnetic spin fluctuation for the $\tau$=1 orbital 
further develops,
and orbital fluctuations are completely suppressed
compared with the well developed antiferromagnetic spin fluctuation.

\begin{figure}[t]
\centerline{\epsfxsize=8.5truecm \epsfbox{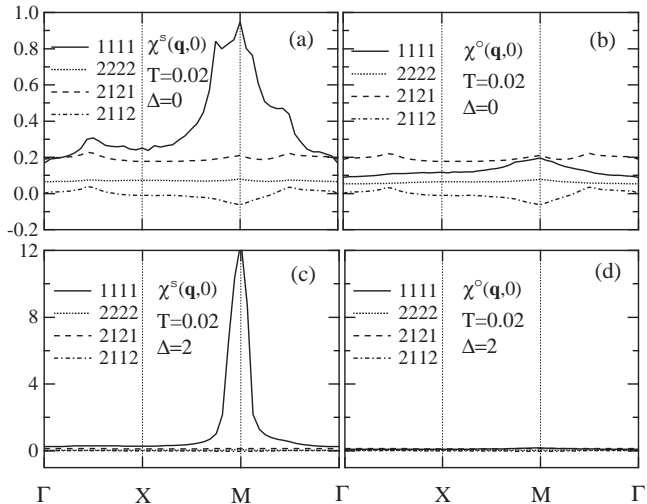} }
\caption{{\bf q}-dependences of 
(a) spin  and (b) orbital susceptibilities for $\Delta$=0. 
(c) and (d) are for $\Delta$=2.
For all figures, solid, dotted, dashed, and dash-dotted lines 
describe components of $\mu\nu mn$=1111, 2222, 2121, and 2112, 
respectively. }
\end{figure}

Let us compare the present results shown in Fig. 3 with those 
obtained within the RPA in the previous work. \cite{manybody}
For $\Delta$=0, 
the antiferromagnetic spin fluctuation in the $\tau$=1 orbital 
already develops without increasing $\Delta$, 
while similar momentum dependences for many components of 
the spin and orbital susceptibilities
has been seen within the RPA. 
When the orbital splitting energy is increased, 
the antiferromagnetic spin fluctuation in the $\tau$=1 orbital 
obtained within the FLEX approximation is considerably enhanced 
in comparison with that within the RPA. 
The considerable development of the antiferromagnetic spin fluctuation 
may rather prevent $d_{x^2-y^2}$-wave superconductivity 
because of decoupling effect 
through the fluctuation exchange self-energy.

\subsection{Phase Diagram}
In Fig.~4, the phase diagram obtained within the FLEX approximation
for the orbital degenerate system is shown, 
where the solid and open circles describe 
the superconducting and antiferromagnetic transition points, respectively.
Although 
we do not carry out the FLEX calculation in the broken symmetry states, 
we may expect that 
the first-order phase transition between 
the $d_{x^2-y^2}$-wave superconducting phase and 
the antiferromagnetic phase takes place in the quasi-two 
dimensional system as discussed for the three-dimensional Hubbard model. 
\cite{Takimoto2}
For the latter case, it has been shown 
that the increase of the three-dimensionality expands (shrinks) 
the magnetic ($d_{x^2-y^2}$-wave superconducting) phase 
and the coexistent phase between the $d_{x^2-y^2}$-wave superconducting 
and magnetic phases appears only for the system having 
the moderate three-dimensionality. 
From this consideration, 
the dotted curve is drawn 
as the first-order phase boundary expected 
between the two phases.

\begin{figure}[t]
\centerline{\epsfxsize=8.5truecm \epsfbox{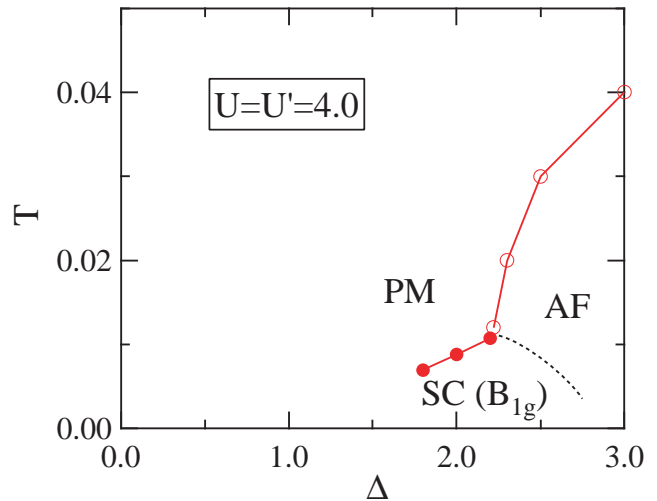} }
\caption{
Phase diagram in the $T$-$\Delta$ plane for $U$=$U'$=4.0 
obtained by the FLEX approximation. 
The dotted curve is drawn by hand 
as expected phase boundary of first-order phase transition 
between the $d_{x^2-y^2}$-wave superconducting phase and 
the antiferromagnetic phase. 
In the figure, PM, AF, and SC(B$_{1g}$) express 
paramagnetic, antiferromagnetic, and 
$d_{x^2-y^2}$-wave superconducting phases, respectively.}
\end{figure}

From Fig.~4, we see that
(1) the spin-singlet superconducting phase 
with $B_{\rm 1g}$-symmetry
appears next to the antiferromagnetic phase and 
(2) $T_{\rm c}$ is enhanced with
increasing the orbital splitting energy $\Delta$.
Recalling the expression of the effective pairing interaction 
for spin-singlet state $T^{\rm s}_{\mu m,n\nu}(q)$, 
the increase of $T_{\rm c}$ with increasing $\Delta$ is ascribed to 
the development of the antiferromagnetic spin fluctuation 
and the suppression of the orbital fluctuation, 
where the spin and orbital fluctuations are destructive each other 
for the spin-singlet pair. 
From these observations, we conclude that the superconducting phase
is induced by the development of the antiferromagnetic spin fluctuation
for the $\tau$=1 orbital 
accompanied by suppression of the orbital fluctuation 
with increasing the orbital splitting energy 
$\Delta$.

\normalsize

\section{Discusion and Summary}

Based on the simple model with orbital degree of freedom, 
we have shown that 
in the system of the density of one electron per site, 
the $d_{x^2-y^2}$-wave superconducting transition temperature 
becomes higher for the larger orbital splitting energy. 
For sufficiently large orbital splitting energy, 
the antiferromagnetic transition takes place 
at a N$\acute{e}$el temperature. 
Here we discuss the experimental results for CeTIn$_5$ 
from the present theoretical results. 
Analyses of experimental data of magnetic susceptibilities 
by using the CEF theory have determined the level schemes 
of CeTIn$_5$ compounds where the two $\Gamma_7$ are lower 
than the $\Gamma_6$. \cite{Takeuchi,Shishido}
The energy splitting between the two $\Gamma_7$ is estimated 
as 68K for CeRhIn$_5$, 61K for CeIrIn$_5$, and 151K for CeCoIn$_5$. 
As we have mentioned in introducing the present model, 
we focus on these orbital splitting energies of the compounds 
regardless of kind of excited Kramers doublet. 
Considering that 
the effect of the orbital degree of freedom is generally quenched 
by increasing the orbital splitting energy, 
the present result may be independent of details of 
level scheme in orbital degenerate models. 
Namely, the present result seems to be 
consistent with the experimental one in the sense 
that $T_{\rm c}$ in CeCoIn$_5$ 
with larger orbital splitting energy 
is higher than that in CeIrIn$_5$. 
Thus, one can expect that 
the orbital splitting energy actually plays 
a role of controlling parameter 
for the superconductivity around the antiferromagnetic phase 
in the heavy fermion system of CeTIn$_5$ compounds. 

On the other hand, the present theory seems to be inadequate 
to explain the antiferromagnetism of CeRhIn$_5$ 
with the level scheme similar to CeIrIn$_5$. 
Recent de Haas-van Alphen experiment of 
CeRhIn$_5$ has reported that the Fermi surfaces are almost 
unchanged up to 2.1 GPa slightly higher than the critical pressure 
at which the curves 
of the superconducting and antiferromagnetic boundaries 
cross each other. \cite{Shishido2}
This may mean that the $f$-electronic states are much lower than 
the Fermi level, and local character of $f$-electron 
is dominant in this compound. 
Thus, it will be a future problem to 
clarify the antiferromagnetism of CeRhIn$_5$ 
by developomg a microscopic theory 
based on the local property of $f$-electrons.

It is instructive to 
recall that the present Hamiltonian 
with considerably large $\Delta$ in the quarter-filling 
is regarded as an effective Hamiltonian in the hole-picture 
for the CuO$_2$ plane of the cuprate La$_2$CuO$_4$ 
which is the parent compound of high-$T_{\rm c}$ cuprate 
where $d_{x^2-y^2}$- and $d_{3z^2-r^2}$-orbitals just correspond to 
$\Gamma_{8}^{(1)}$ and $\Gamma_{8}^{(2)}$ in the present model, 
respectively. 
Considering this similarity, it may be a reasonable speculation 
that a superconducting phase may appear 
next to the antiferromagnetic phase 
of La$_2$CuO$_4$ in some situation 
where distance between the apical O- and the planer Cu-sites 
is shrunk, the orbital splitting energy 
between $d_{x^2-y^2}$- and $d_{3z^2-r^2}$-orbitals is decreased, 
after all.

In summary, based on the effective microscopic model with
orbital degeneracy for $f$-electron systems,
we have developed the strong-coupling theory for superconductivity. 
Considering the time reversal symmetry involved in the model, 
one can show that the orbital antisymmetric component of gap function 
hardly contributes to superconductivity because of the odd-frequency 
dependence. 
Using the FLEX approximation in which 
the spectra of $f$-electrons, the spin and orbital fluctuations, 
and the superconducting gap functions are determined consistently, 
it has been shown that the $d_{x^2-y^2}$-wave 
superconducting phase is induced by increasing the orbital 
splitting energy which leads to the development and suppression 
of the spin and orbital fluctuations, respectively. 
Based on these results, 
we have proposed that the orbital splitting energy is
the key parameter controlling the changes from the paramagnetic
to the antiferromagnetic phases with 
the $d_{x^2-y^2}$-wave superconducting phase in between.


\section*{Acknowledgement}

The authors would like to thank T. Maehira 
for many valuable discussions.
T. H. and K.U. are separately supported 
by the Grant-in-Aid for Scientific Research 
from Japan Society for the Promotion of Science.

\vskip-0.5cm


\begin{thebibliography}{99}
\bibitem{footnote1}
Although orbital ordering states are affected by 
kind of excited Kramers doublet, 
such states are not considered in this paper.
\end{thebibliography}
\end{document}